\documentclass{ws-p8-50x6-00}

\def\bea{\begin{eqnarray}}
\def\eea{\end{eqnarray}}

\def\l{\lambda}

\def\0{0_{\perp}}

\begin{document}

\title{
{\normalsize DESY 01-185 \hfill\mbox{ }}\\[5ex]
Physics at HERA II}

\author{W. Buchm\"uller}

\address{Deutsches Elektronen-Synchrotron DESY, 22603 Hamburg, Germany 
\\E-mail: buchmuwi@mail.desy.de}


\maketitle

\abstracts{
Deep inelastic electron proton scattering at HERA II will allow precise
studies of QCD and stringent tests of physics beyond the standard model. We 
discuss these two aspects of DIS with emphasis on the regime of high gluon 
densities at small $x$ and on scalar quark production in supersymmetric 
theories with broken R-parity.}

\section{Introduction}

In deep inelastic scattering at HERA one studies the interactions of
electrons with quarks in a wide range of momentum transfers $Q^2$ and
center-of-mass energies $\sqrt{\hat{s}}= \sqrt{xs}$, up to 
$\sqrt{\hat{s}}_{max} = 318$~GeV. The `Virtues of HERA' have been
identified long ago\cite{mai83}. DIS at small $Q^2$ probes the structure
of the proton and allows a variety of QCD tests. DIS at large $Q^2$ is
sensitive to electroweak interactions, to a possible structure of quarks 
and electrons and, last but not least, to new particles and interactions 
predicted by extensions of the standard model. 

HERA I has made important contributions to all of these topics. The increase
of luminosity by a factor of 10 and the availability of longitudinal 
polarization for electrons and positrons at HERA II will widen the
physics scope substantially. In the following we
shall illustrate this with some examples concerning strong interactions,
electroweak interactions and physics beyond the standard model. This
complements the reports of the H1\cite{els01} and ZEUS\cite{fos01} 
collaborations on their physics program at HERA II. Detailed discussions
can be found in the proceedings of previous HERA workshops\cite{hera87} 
as well as in the review articles \cite{ac99,wol01}.

\section{Beyond the Standard Model}

In connection with the Higgs mechanism of mass generation for vector bosons
and fermions new physics beyond the standard model is expected at energies
${\cal O}$(1~TeV). There are two classes of extensions of the standard model.
In the first class a revolutionary change is predicted, either as quark-lepton
compositeness or as manifestations of large extra dimensions. One then expects
towers of excitations of the known elementary particles, either excited
quarks and leptons or Kaluza-Klein excitations related to the TeV string 
scale. In the second class of extensions only a `mild' modification of the
standard model is predicted, the occurence of supersymmetry, where each
particle aquires a superpartner. These extensions successfully predict the
unification of all interactions at the GUT scale ${\cal O}(10^{16}$~GeV).
Via the seesaw mechanism they can also naturally account for the small
neutrino masses indicated by the solar and atmospheric neutrino deficits.

\begin{figure}
\vspace*{-.5cm}
\centerline{
\epsfig{figure=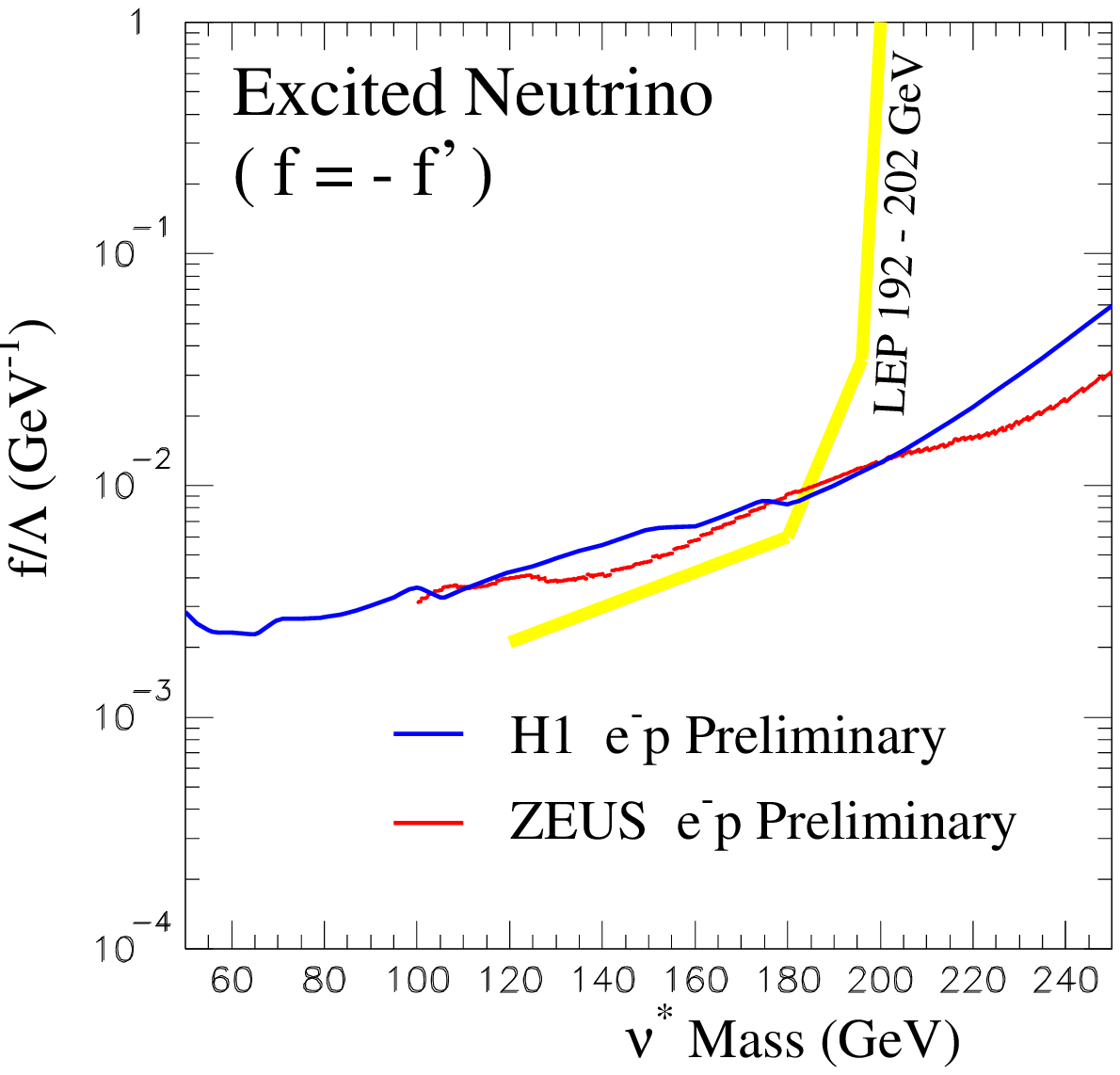,width=6cm}
\epsfig{figure=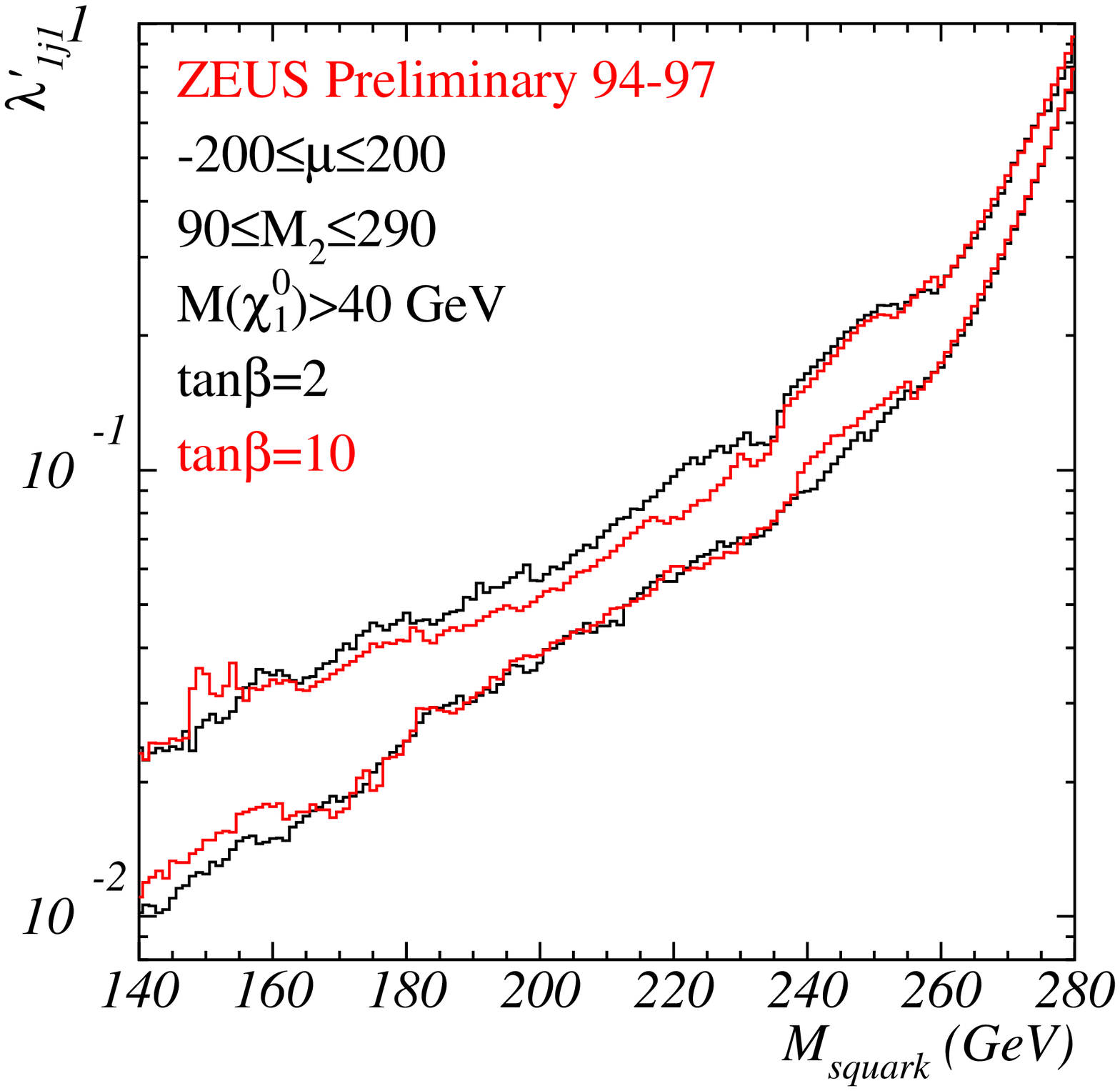,width=5.5cm}}
\caption{Bounds on coupling versus mass for excited 
neutrinos\protect\cite{exneu} (left), and  
scalar quarks\protect\cite{zrp} (right).}
\label{rpar}
\end{figure}

HERA is particularly sensitive to those new particles for which single
production is possible in ep-collisions. These include excited quarks
and leptons, leptoquarks and scalar quarks in supersymmetric models with
broken R-parity. For excited neutrinos the LEP limit of 200~GeV could be
significantly extended as shown on the left in fig.~(\ref{rpar}). 
On the right the range of upper limits on couplings of scalar quarks is 
shown as function of the squark mass for a class of supersymmetric models.  
The Yukawa couplings determine the production cross section, e.g.
$\sigma(e_R^+ d_R\rightarrow \widetilde{u}_i) \propto {\lambda'}_{1i1}^2$, with
$\widetilde{u}_1 = \widetilde{u}$, $\widetilde{u}_2 = \widetilde{c}$, 
$\widetilde{u}_3 = \widetilde{t}$. 

\begin{figure}
\vspace*{-.5cm}
\centerline{
\epsfig{figure=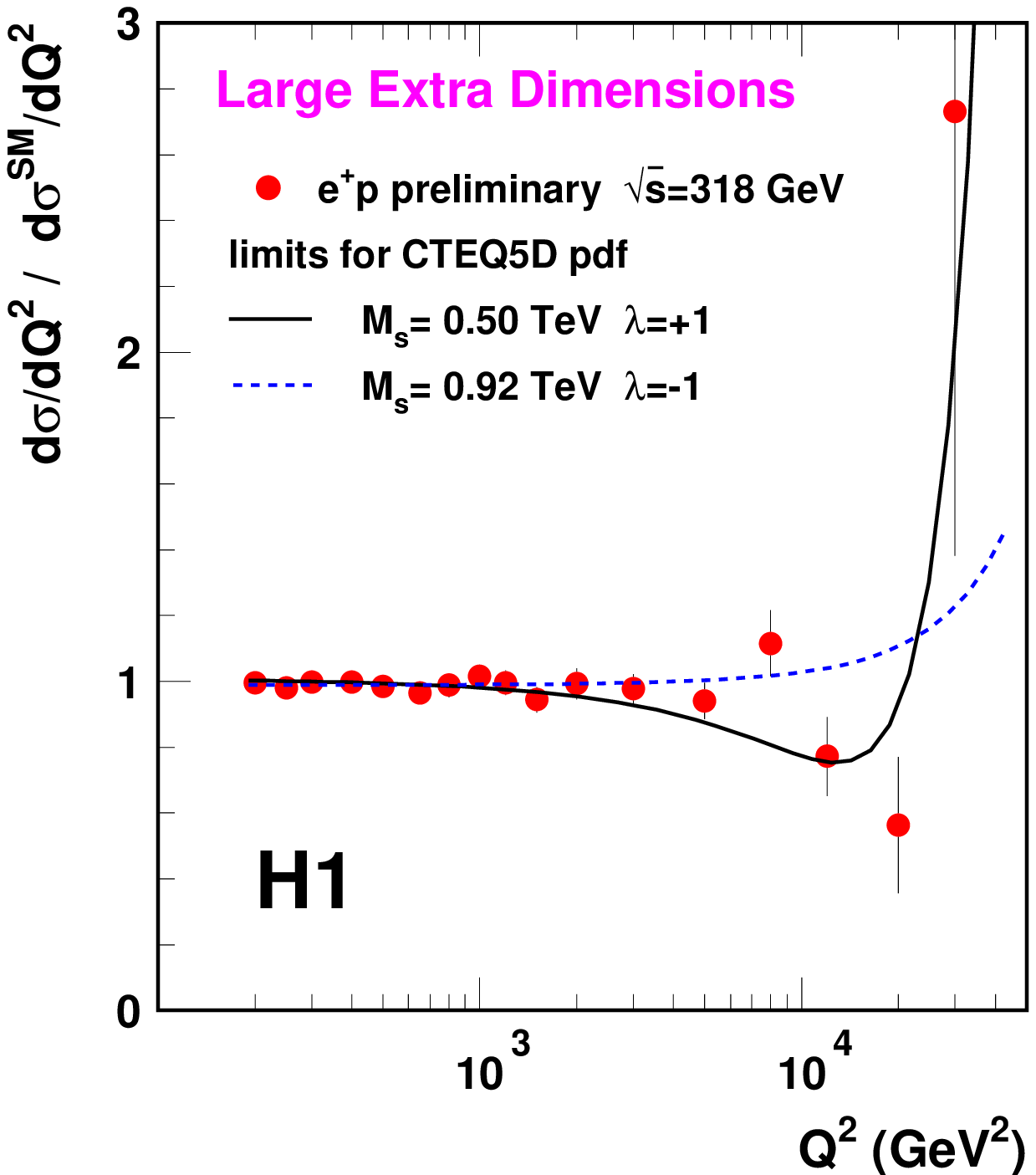,width=6cm}
\epsfig{figure=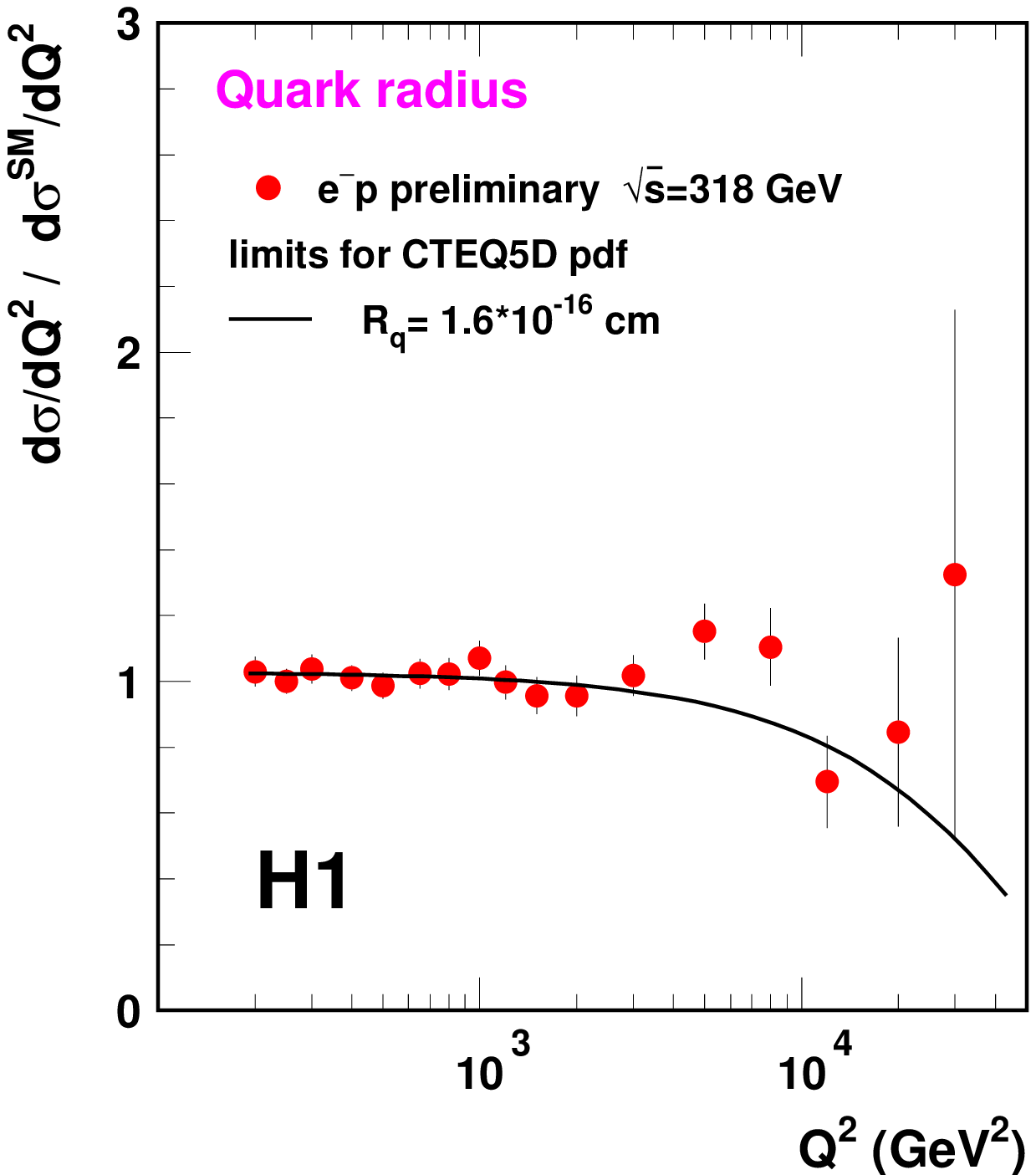,width=6cm}}
\caption{Effect of large extra dimensions of size $\sim 1/M_s$ (left) and of a 
finite quark radius $R_q$ (right) on the neutral current DIS cross section. 
From ref.\protect\cite{extrah1}.}\label{extra}
\end{figure}

Heavier particles beyond the kinematic reach of HERA lead to effective
four-fermion interactions\cite{kuz01} with strength $4\pi/\Lambda^2$. 
Depending on the chirality structure lower bounds on $\Lambda$ up to 9~TeV 
have been obtained. For the vector-current coupling the bound on $\Lambda$ can
be interpreted as an upper bound on the electromagnetic quark radius.
The present limit is $R_q = 1.6\times 10^{-16}$~cm (cf.~fig.~(\ref{extra})).
Contact interactions are also induced by the Kaluza-Klein tower of gravitons
in theories with large extra dimensions. The present bound on the 
corresponding mass scale is about 1~TeV  (cf.~fig.~(\ref{extra})). At
HERA II all these bounds can be improved by about a factor of three.

Supersymmetric theories with broken R-parity are for HERA the most promising
extensions of the standard model\cite{psd96}. In a class of these models
Majorana neutrino masses are generated radiatively. It is then possible to
relate the neutrino mixing matrix to squark production cross sections in
ep-collisions\cite{bw01}. The analyses of the solar and atmospheric neutrino
anomalies favour large neutrino mixings and therefore a neutrino mass matrix
of the form
\bea
m_{\nu ij} = m f_{ij}\;,
\eea
where $m < 1$~eV and $f_{ij} = {\cal O}(1)$ for all $i,j=1\ldots 3$. As we
shall see, the large mixings among the neutrinos can lead to large couplings of
electrons to new particles. The small value of the neutrino mass scale $m$ 
can be generated radiatively or by mass mixing via the seesaw mechanism. 

In the supersymmetric standard model Yukawa interactions are described by the
superpotential
\bea\label{neumass}
W = h_{eij} E^c_i L_j H_1 +  h_{dij} Q_i D^c_j H_1 +
    h_{uij} Q_i U^c_j H_2\;,
\eea
where $H_1$ and $H_2$ are two Higgs doublets with vacuum expectation
values $v_i=\langle H_i \rangle$, $i=1,2$, and $v_2/v_1 = \tan{\beta}$.
The fact that the lepton
doublets $L_i$ and the Higgs doublet $H_1$ have the same hypercharge,
and therefore identical gauge quantum numbers, motivates the introduction
of an additional R-parity violating part of the superpotential,
\bea\label{superR}
W_R =  {1\over 2}\l_{ijk} L_i L_j E^c_k +  \l'_{ijk} L_i Q_j D^c_k\;.
\eea
The couplings $\l$ and $\l'$ are in principle arbitrary. However, the
same reason that leads to the introduction of these R-parity violating 
couplings also
suggests the following connection between $\l$, $\l'$ and the Yukawa couplings
$h_e$ and $h_d$,
\bea\label{connect}
\l_{ijk} = \l_i h^T_{ejk}+\l_j h^T_{eik}\;, \quad  \l'_{ijk} = \l'_i h_{djk}\;.
\eea
$W_R$ is then obtained from $W$ by a rotation among the fields $(L_i,H_1)$.
Such a `flavour alignment' suppresses the rates of flavour changing processes
in the down quark sector ($\Delta S = 1,2, \Delta B = 1,2$). 

$W_R$ violates lepton number\cite{glt01}. Hence, Majorana neutrino masses 
are induced,
\bea
m_{\nu ij} = \l'_i\l'_j m_{\nu}^{(d)} + \l_i\l_j m_{\nu}^{(e)}\;,
\eea
where $m_{\nu}^{(d,e)}$ depends on the Yukawa couplings $h_{d,e}$ and the
soft supersymmetry breaking parameters. In order to obtain the neutrino
mass matrix (\ref{neumass}) one needs $\l_i,\l'_j = {\cal O}(1)$. The choice 
$\l_i=\l'_j=1$ leads to the `democratic' mixing matrix.

Predictions of this model with R-parity breaking are certain flavour changing
processes, e.g. $BR(D^0 \rightarrow \mu\mu,\mu e) \sim 10^{-5}$ and, in
particular, the couplings for squark production, for instance,
\bea
\l'_{111} \sim {m_d\over v} \tan{\beta} \sim 0.003\; , 
\quad \l'_{122} \sim {m_s\over v} \tan{\beta} \sim 0.05\;.
\eea
The value of $\l'_{111}$ is consistent with the upper bound from neutrinoless 
double beta-decay. The present upper bound\cite{h1rp} $\l'_{122} < 0.29$ based
on 37~pb$^{-1}$ suggests that the sensitity of $\l'_{122} \sim 0.05$ will
be reached at HERA II.

\section{Electroweak Interactions}

DIS at large $Q^2 \sim 10^4$~GeV$^2$ has so far tested electroweak 
unification, i.e., the approximate equality of neutral current and
charged current cross section, $\sigma(NC) \sim \sigma(CC)$. At HERA II
polarization will allow to test the classic prediction of the electroweak
theory,
\bea
\sigma^{CC}(e^-_R p) = \sigma^{CC}(e^+_L p) = 0 \;.
\eea 
The exchange of right-handed $W_R$-bosons leads to a non-zero cross section.
One expects a sensitivity to masses $m_{W_R} \simeq 600\ldots 800$~GeV,
which corresponds to the present bounds from direct production and electroweak
precision tests\cite{rpp00}.
\begin{figure}
\begin{center}
\vspace*{-.5cm}
\epsfig{figure=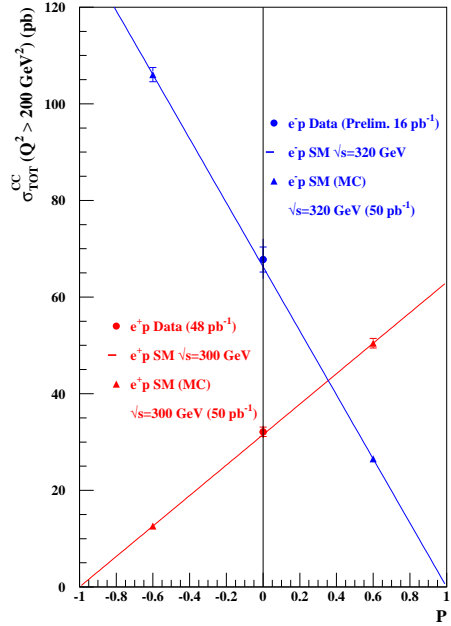,width=6cm}
\label{pol}
\end{center}
\caption{Charged current cross section as function of polarization $P$. 
Preliminary ZEUS data for $P=0$ and Monte Carlo simulations for $P\neq 0$.
From ref.\protect\cite{fus01}.}
\end{figure}
The expected precision for the charged current cross section obtained after
one year of running is shown in fig.~(\ref{pol}) as function of the degree of 
polarisation $P$.

The weak mixing angle $\sin^2(\theta_W)(M_Z)$ can be determined from a
measurement of the polarization asymmetry
\bea
A(e_L^- - e_R^-) = 
{d\sigma(e_L^-)-d\sigma(e_R^-)\over d\sigma(e_L^-)+d\sigma(e_R^-)}\;.
\eea 
For an integrated luminosity of 500~pb$^{-1}$ one expects an error of about
$1\%$\cite{hai01}. This is less accurate, but complementary to the LEP and
SLC measurements.

Polarization is also crucial to determine the properties of discovered new
particles. For instance, for the scalar quarks discussed in the previous 
section one has
\bea
\sigma(e_R^+ p \rightarrow \widetilde{c} X) \propto {\l'}_{122}^2\ ,\quad
\sigma(e_L^+ p \rightarrow \widetilde{c} X) = 0\; .
\eea
Such a measurement would prove that the discovered new scalar colour-triplet
particle is the superpartner of a left-handed quark.

\section{Strong Interactions}

In deep inelastic scattering at HERA I many features of QCD have already been 
studied in great detail. These include 
\begin{itemize}

\item proton and photon structure functions,

\item jets and event shapes, 

\item determination of $\alpha_s$,

\item hadronic final states,

\item instanton induced processes,

\item production of charm and bottom,

\item vector meson production,

\item diffractive processes.

\end{itemize}
All these quantities and processes will be studied with higher precision at 
HERA II, and in particular the search for instanton induced 
processes\cite{rs01} will be significantly improved.

In the following I shall concentrate on those aspects which
have been most intriguing at HERA I. These are

\begin{enumerate}

\item the rapid rise of parton densities at small $x$\cite{h1zrise}, and

\item the large fraction of diffractive events at small $x$\cite{h1zdiff} .

\end{enumerate}

The rise of the structure functions at small $x$ has been anticipated
based on the QCD renormalization group equations for moments\cite{rx74} 
and the corresponding DGLAP evolution 
equations\cite{dglap} for parton densities in the GRV model\cite{grv93}. 
In addition, at small $x$ large logarithms $\sim \alpha_s \ln{1/x}$ 
become important and have to be resummed. This is achieved
by means of the BFKL equation\cite{bfkl} which leads to the prediction
of a power-like growth, $F_2 \propto x^{-\lambda}$. The rapid rise
at small $x$ reflects the strong radiation of gluons at high energies
leading to large gluon densities. However, it still remains to be understood 
to what extent this rise can be described perturbatively and where it
reflects non-perturbative effects, i.e., screening corrections\cite{grv83} 
and input parton distributions.

\begin{figure}
\begin{center}
\vspace*{-.5cm}
\epsfig{figure=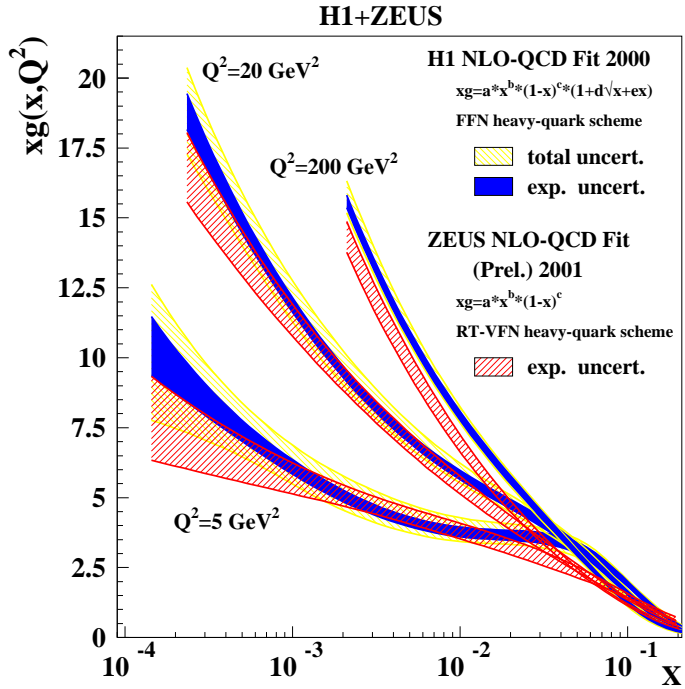,width=9cm}
\end{center}
\caption{Gluon distribution function determined by a NLO-QCD fit to the
structure function $F_2(x,Q^2)$ based on H1 data\protect\cite{wal01} and ZEUS 
data\protect\cite{nag01}, respectively.
\label{gluon}}
\end{figure}

\begin{figure}
\vspace*{-.5cm}
\centerline{
\epsfig{figure=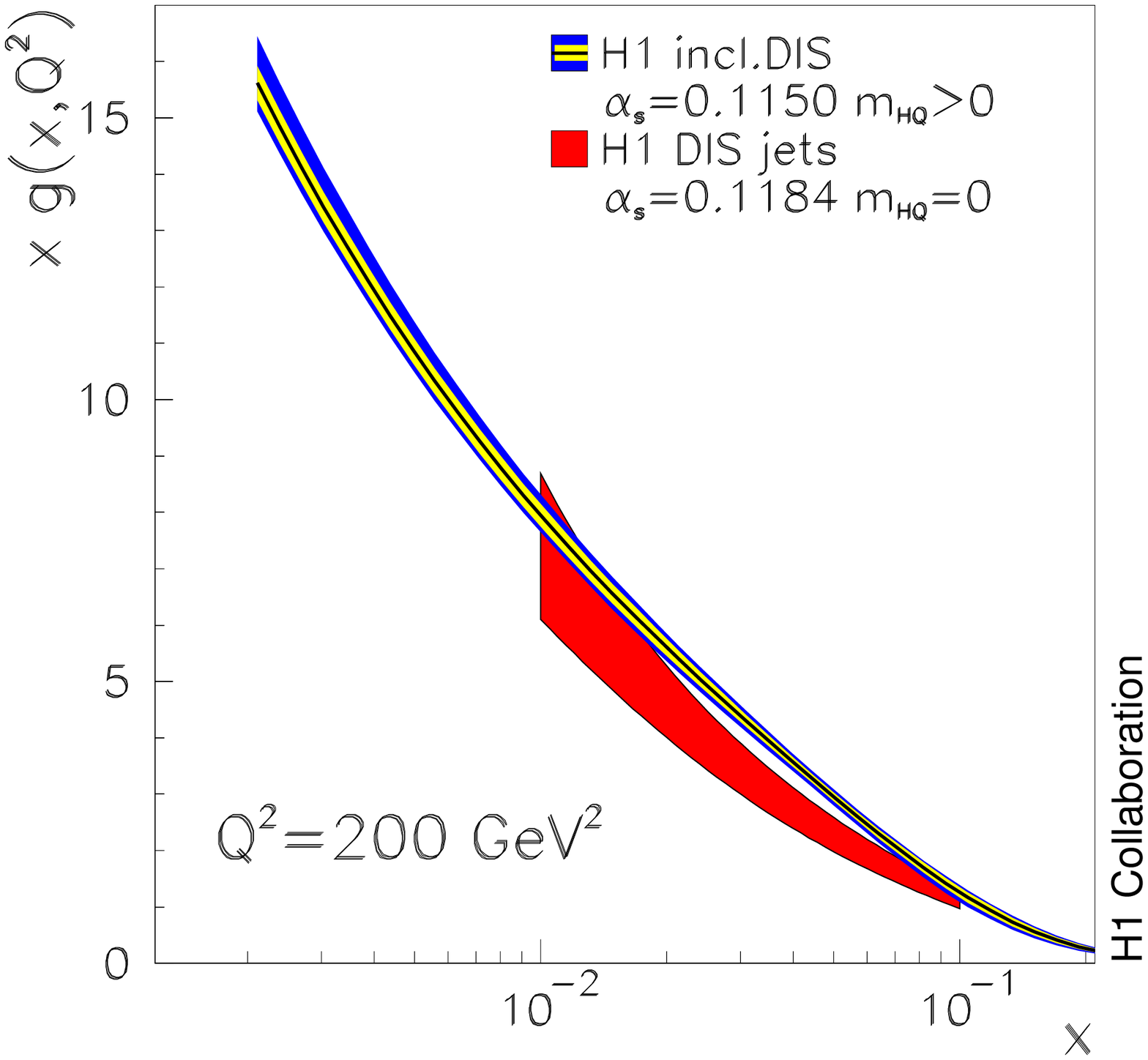,width=5.5cm}
\epsfig{figure=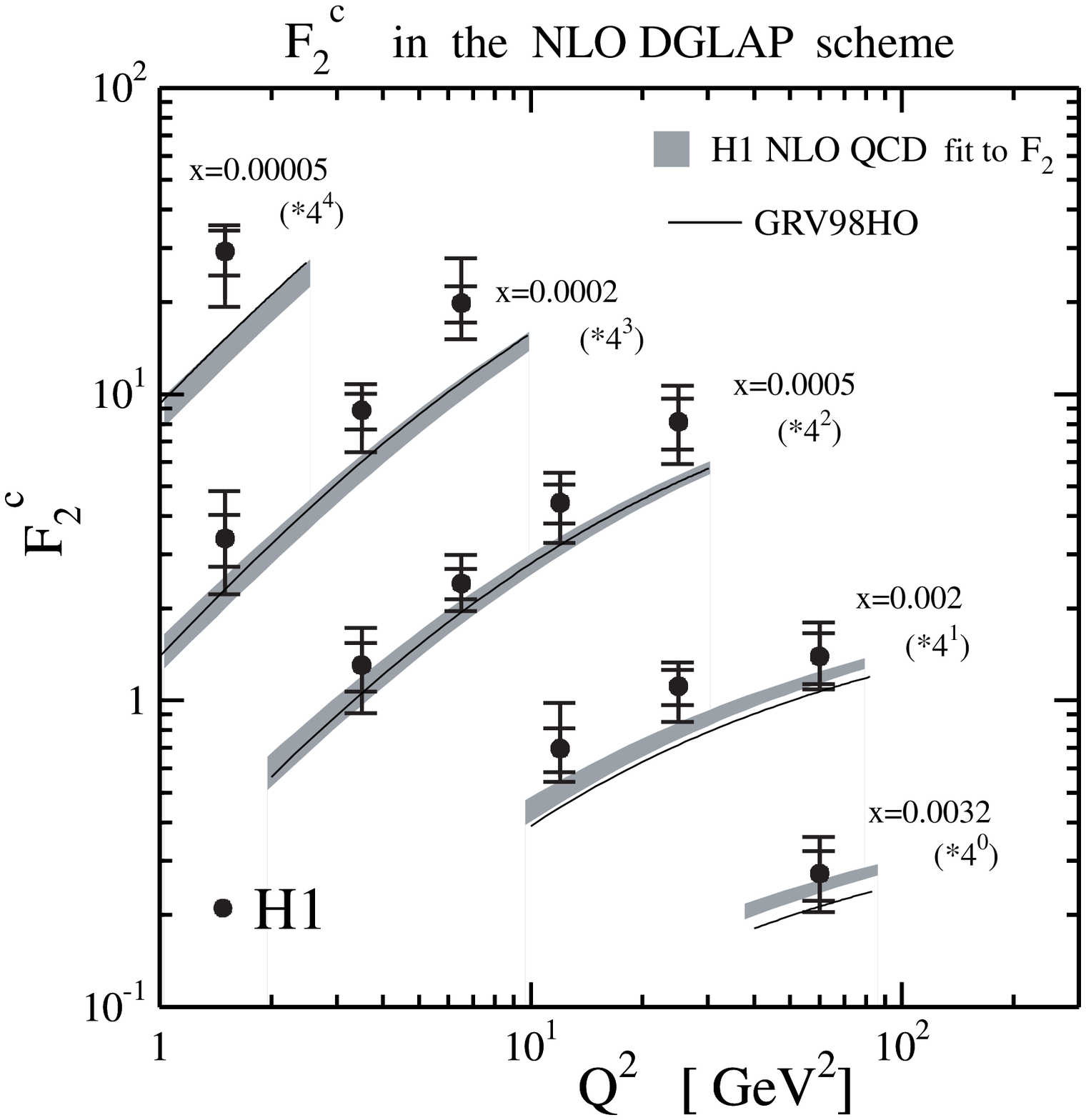,width=6cm}}
\caption{Gluon distribution function extracted from a NLO-QCD fit to dijet 
rates (left); the charm contribution to $F_2$ (right). \label{jcgluon}}
\end{figure}

In order to identify possible non-perturbative effects, it is important to 
check how accurately the DGLAP approach describes different processes.
A particularly interesting quantity is the gluon density which can be
extracted from scaling violations of the structure functions,
\begin{equation}
xg(x,Q^2) \propto \frac{\partial F_2(x,Q^2)}{\partial\ln Q^2}\;. \label{ctrans}
\end{equation}
The gluon density extracted in a NLO analysis from the H1 and ZEUS data 
is shown in fig.~(\ref{gluon}) for different values of $Q^2$. The related
determination of $\alpha_s$ has by now reached the remarkable 
precision\cite{alh1,alz},
\bea
\alpha_s(M_Z^2) &=& 0.1150 \pm 0.0017 (exp) \pm 0.005 (theory)\ (H1),\\
\alpha_s(M_Z^2) &=& 0.117 \pm 0.001 (stat + uncorr) 
                               \pm 0.005 (corr) \ (ZEUS, prel.).
\eea
Note, that the total error of the H1 result is dominated by the theoretical 
uncertainty.

Contrary to structure functions, the gluon
density enters at leading order in the dijet cross section and in $F^c_2$, 
the charm contribution to the structure function $F_2$. In fig.~(\ref{jcgluon})
the NLO gluon density extracted from dijets is compared with the one
determined from scaling violations; further, the corresponding prediction
for $F^c_2$ is compared with data.
It is clear that the higher accuracy at HERA II will lead to a stringent
test of the DGLAP framework, which will require NNLO theoretical 
calculations for structure functions and jet cross sections.

\begin{figure}
\begin{center}
\vspace*{-.5cm}
\epsfig{figure=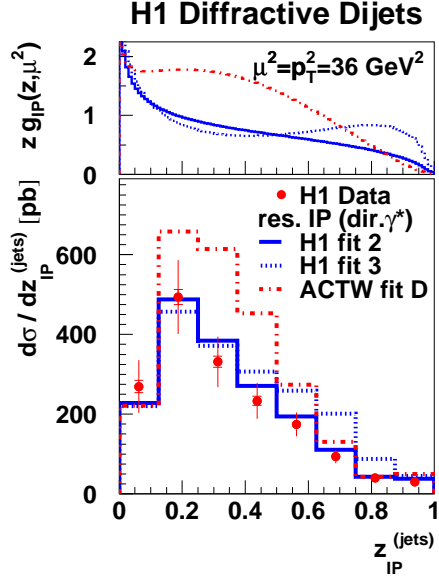,width=6cm}
\end{center}
\caption{Diffractive dijet cross section as function of 
$z_P^{(jets)} ( = \beta)$,
compared with predictions based on different diffractive gluon densities.
The corresponding gluon densities $g(z,\mu^2)$ are shown in the upper pannel.
From ref.\protect\cite{h101}.}
\label{h1dij}
\end{figure}

A puzzling phenomenon in DIS at small $x$ is the occurence of a large
fraction ${\cal O} (10 \%)$ of diffractive events\cite{h1zdiff}. 
This came as a big surprise for most experts in QCD, although it had
been anticipated based on Regge theory\cite{dl87}. However, the large rapidity
gap events in DIS are difficult to understand in the parton picture 
to which almost everybody became used during the past 25 years because of the 
successes of perturbative QCD.

We now have learned that, like inclusive structure functions, also diffractive 
structure functions
can be expressed as convolution of parton cross sections with diffractive
parton densities\cite{tv94,bs94}. This factorization has been proven 
for inclusive and diffractive DIS with corresponding rigour\cite{col98}.
The variation of the leading twist structure function with 
$Q^2$ is given by
\begin{eqnarray}
\lefteqn{ Q^2 {\partial\over \partial Q^2} F_2^D(\xi,\beta,Q^2) =} \\
&&2\sum_q e_q^2x {\alpha_s\over 2\pi}
\int_{\beta}^1 {db\over b} 
\left(P_{qq}\left({\beta\over b}\right){dq(b,\xi,\mu^2)\over d\xi}
+ P_{qg}\left({\beta\over b}\right){dg(b,\xi,\mu^2)\over d\xi}\right)\;.
\nonumber\label{ftd}
\end{eqnarray}
Here $\xi \equiv x_P$ is the fraction of momentum lost by the proton, and 
$\beta=Q^2/(Q^2+M^2)$ where $M$ is the diffractive mass, $M^2=(q+\xi P)^2$.
For comparison, in inclusice DIS, $x=Q^2/(Q^2+W^2)$ where $W^2=(q+P)^2$
is the total invariant mass squared of the complete hadronic system.  
$dq/d\xi$ and $dg/d\xi$ are the diffractive quark and gluon densities,
respectively, and $P_{qq}$ and $P_{qg}$ are the usual splitting functions.
Note, that the $Q^2$-evolution affects only the $\beta$-dependence of the 
structure function and not the $\xi$-dependence. Hence, the $\xi$-dependence 
is an entirely non-perturbative property of the proton. It corresponds 
to the dependence on the total hadronic energy $W$ for fixed $\beta$, since
$W^2 \simeq Q^2/x = Q^2/(\xi\beta)$. 
In models with Regge factorization\cite{is85} the $\xi$-dependence is 
given by Regge theory. For the diffractive gluon density one has
$dg(\xi,\beta,Q^2)/d\xi \propto \xi^{1-2\alpha_P(0)} g(\beta,Q^2)$.

\begin{figure}
\begin{center}
\epsfig{figure=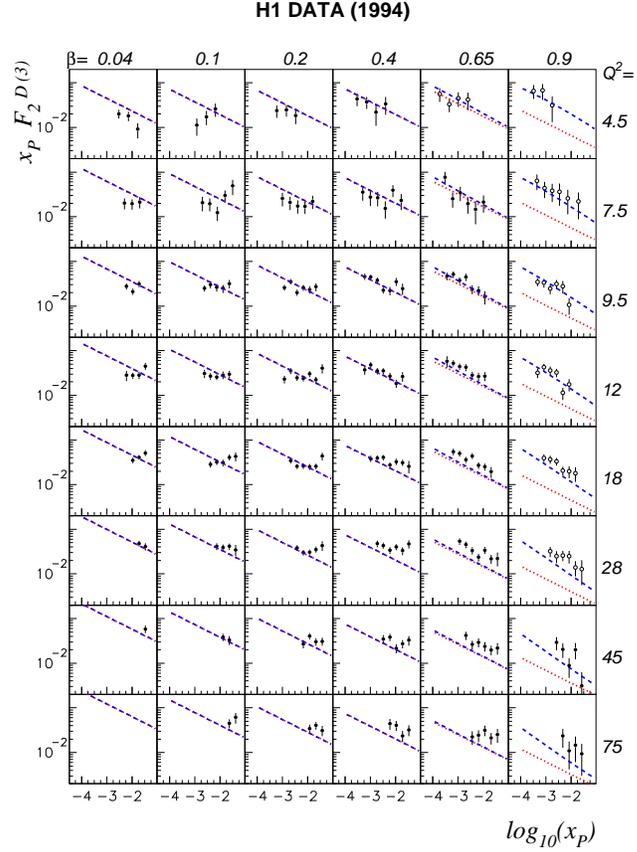,width=9cm}
\end{center}
\caption{Prediction for the diffractive structure function $x_P F_2^{D(3)}$ 
compared with H1 data. The dashed lines correspond to the leading twist
contribution with the twist-4 component added. The leading twist contribution
is shown by the dotted lines. From ref.\protect\cite{gbw01}.\label{f2dgbw}}
\end{figure}

Diffractive DIS can be analyzed in analogy to inclusive DIS\cite{gs96}.
The diffractive structure function yields the diffractive quark
distribution. Its scaling violation, and also diffractive
dijet and charm production determine the diffractive gluon density.
In fig.~(\ref{h1dij}) the measured dijets rates (lower panel) are compared 
with predictions based on different diffractive gluon densities $g(z,\mu^2)$
(upper panel) which have been extracted from fits to the diffractive 
structure function. At present quantitative tests
are just beginning. From HERA II we can expect a precise determination of
the diffractive gluon density. Here also higher twist effects have to be taken
into account at large $\beta$\cite{bex99,ht01}. The results of a recent 
analysis including such effects are shown in fig.~(\ref{f2dgbw}).  

\begin{figure}
\begin{center}
\epsfig{figure=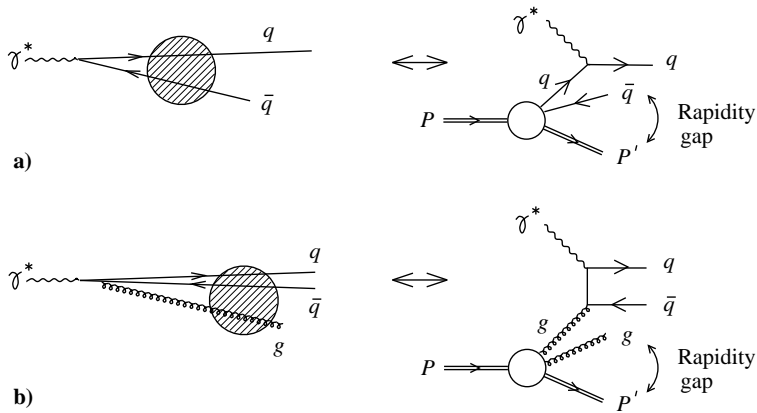,width=10cm}
\end{center}
\caption{Diffractive DIS in the proton rest frame (left) and 
the Breit frame (right); asymmetric quark fluctuations correspond to 
diffractive quark scattering, asymmetric gluon fluctuations to diffractive 
boson-gluon fusion. \label{f2d}}
\end{figure}

What do we learn from a determination of diffractive quark and gluon densities?
A comparison of inclusive and diffractive DIS is particularly interesting
in the proton rest frame\cite{bjo72}. Here diffractive and non-diffractive  
processes can be understood as scattering of partonic fluctuations of the 
photon on the proton. The diffractive quark and gluon densities then
correspond to asymmetric quark and gluon fluctuations (cf.~fig.~(\ref{f2d}))
projected onto the colour singlet state.
The formation of the final state is a non-perturbative phenomenon which 
depends on soft momenta and the properties of confinement. Hence, a comparison
of diffractive and non-diffractive processes should help to understand 
non-perturbative properties of the proton. Correspondingly, one expects that 
at small $x$ diffractive and non-diffractive processes have a similar 
dependence on the total hadronic energy $W$\cite{bh96}.

\begin{figure}[t]
\begin{center}
\parbox{10.5cm}{~\epsfig{file=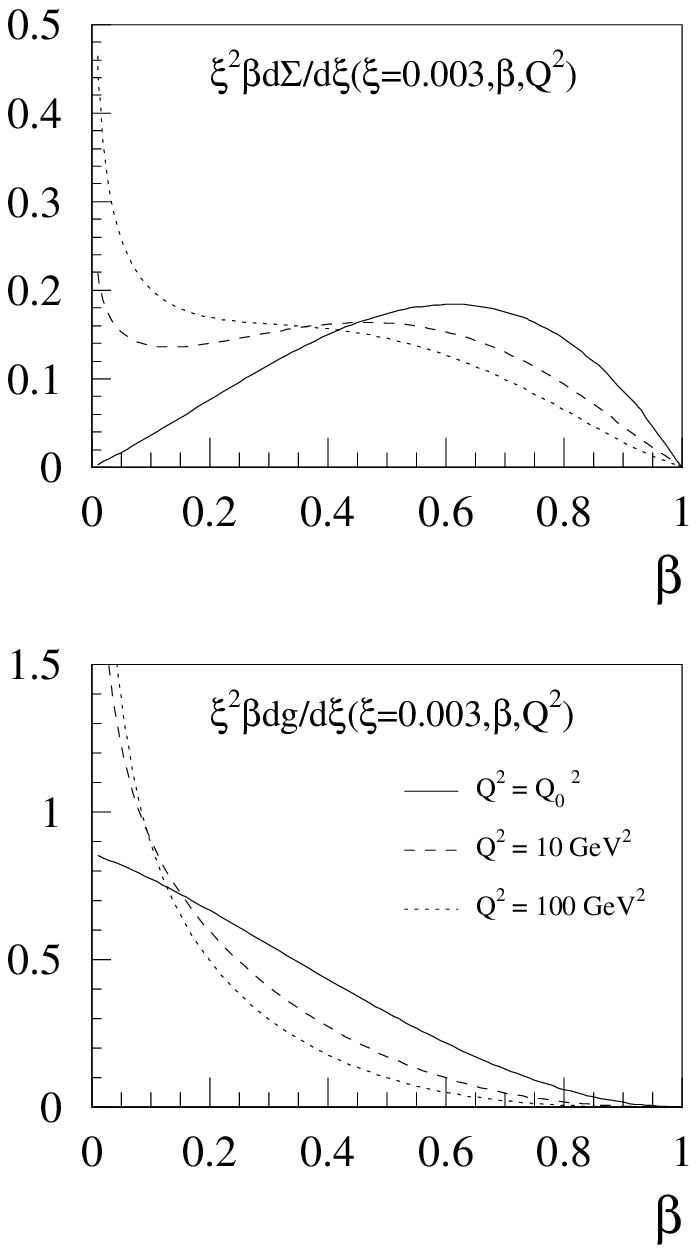,
width=5cm}\hspace{0.5cm}~\epsfig{file=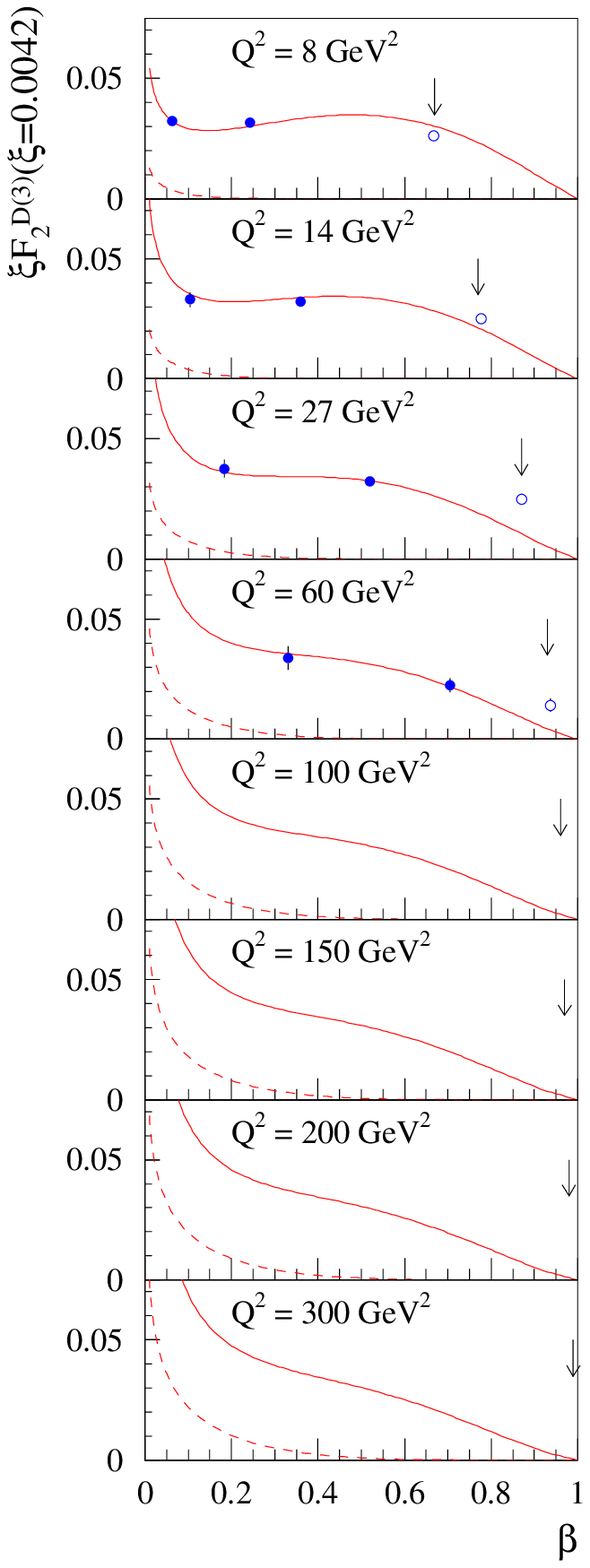,width=5cm}}
\end{center}
\caption{Diffractive singlet quark and gluon distributions at an initial
scale $Q_0^2$ and after $Q^2$ evolution (left); diffractive structure function
$F_2^{D(3)}$ for different values of $Q^2$ compared with ZEUS 
data\protect\cite{zeus99}. The arrows indicate the value of $\beta$ beyond 
which higher twist terms are expected to become important.  
From ref.\protect\cite{bgh99}.\label{dqgd} }
\end{figure}

An interesting toy model for a comparison of inclusive and diffractive DIS 
is a large hadronic target\cite{mlv94,hw98}. In this model the cross section 
for a dipole of transverse size $y$ is of the Glauber-type\cite{nz91}
\begin{equation}
\sigma(y) = \sigma_0 \left(1 - e^{-ay^2}\right)\;.
\end{equation}
The corresponding diffractive\cite{bgh99,gbw01} and 
inclusive\cite{bgh99,mue99} parton densities have been explicitly calculated.
The model can also be used to estimate saturation and higher twist 
effects\cite{gbw99}. 

For the diffractive quark and gluon densities one obtains\cite{bgh99} 
\bea
\xi^2\frac{dq\left(\beta,\xi,Q_0^2\right)}{d\xi}  =  \frac{a\Omega N_c 
(1-\beta)}{2\pi^3} f_q(\beta)\; , \label{qdif} \\
\xi^2\frac{dg\left(\beta,\xi,Q_0^2\right)}{d\xi}  =  \frac{a\Omega N_c^2 
(1-\beta)^2}{2\pi^3\beta} f_g(\beta)\; . \label{gdif} 
\eea
Here $\Omega$ is the transverse size of the target, $N_c$ is the number of
colours, and $Q_0$ is the scale where the model calculation is matched
to the perturbative evolution. A non-trivial dependence of the parton 
densities on $\xi = x_P$ may be introduced
by assuming a $\xi$-dependence either of the transverse size\cite{bgh99},
i.e., 
$\Omega = \Omega(\xi)$, or of the saturation scale\cite{gbw01} $a=a(\xi)$,
leading to the same phenomenology.
Note, that the functions $f_q(\beta)$ and $f_g(\beta)$ are parameter free
predictions of the model.
The singlet quark density $d\Sigma/d\xi = 6 dq/d\xi$ and the
gluon density $dg/d\xi$ are shown in fig.~(\ref{dqgd}). Their 
$\beta$-dependence is an interesting non-perturbative property of the proton.  

Fig.~(\ref{dqgd}) also shows a comparison between a theoretical 
prediction\cite{bgh99} for the diffractive structure function and the presently
available ZEUS data\cite{zeus99}. 
Further data from HERA I and HERA II will allow an
extension to larger values of $Q^2$ where theory predicts almost no change.
It will be important to check whether the DGLAP approach is indeed
correct for the diffractive structure function. Further, because of
the high statistics at HERA II it will be possible to determine the diffractive
gluon density also from dijet and charm production. One can then study the
non-perturbative dependence on $\xi$ and the normalization relative 
to the inclusive gluon density. Particularly interesting will be 
the comparison of event shapes in inclusive and diffractive DIS. In this way
we will get for the first time some insight into the regime of high gluon
densities in QCD\cite{kwx97}.

\section{Summary}

The increase in luminosity at HERA II will allow  precise measurements
of various observables and processes in QCD. Particularly interesting is the
determination of $\alpha_s$, where already now the precision of LEP has
almost been reached. The comparison of the measured running coupling 
$\alpha_s(\mu^2)$ with lattice calculations will lead to an important 
quantitative test of QCD. Further, the
search for instantons can be improved, and one can test 
the DGALP framework by comparing determinations of the gluon density
from different processes. In this way, HERA can probe the regime of
high gluon densities in a clean and controlable way.

The high luminosity and polarization will also lead to new tests of the 
electroweak theory. Searches for physics beyond the standard model will be
significantly improved, in particular with respect to scalar quark production
in models with R-parity breaking. Clearly, this sensitivity will also 
allow the discovery of phenomena not anticipated in this contribution.

\section*{Acknowledgments}

I would like to thank T.~Gehrmann, K.~Golec-Biernat, D.~Haidt, M.~Klein,
M.~Kuze, P.~Schleper and G.~Wolf for their help in the preparation of
this contribution and the organizers of DIS2001 for the splendid hospitality 
in Bologna.


\begin{thebibliography}{99}

\bibitem{mai83}
L.~Maiani, {\it The Virtues of HERA}, in Proc. {\it Experimentation at HERA},
Amsterdam 1983, DESY HERA 83/20

\bibitem{els01}
E.~Elsen, these proceedings

\bibitem{fos01}
B.~Foster, these proceedings

\bibitem{hera87}
Proc.~{\it The HERA Workshop}, Vol.~I,II, Hamburg 1987, ed. R.~D.~Peccei;\\
Proc.~{\it Physics at HERA}, Vol.~I,II,III, Hamburg 1991, 
eds. W.~Buchm\"uller, G.~Ingelman;\\
Proc.~{\it Future Physics at HERA}, Hamburg 1996, eds. G.~Ingelman, 
A.~De Roeck, R.~Klanner

\bibitem{ac99}
H.~Abramowicz, A.~Caldwell, Rev.~Mod.~Phys. {\bf 71} (1999) 1275

\bibitem{wol01}
G.~Wolf, hep-ex/0105055

\bibitem{exneu}
H1 Collaboration, Paper 956 submitted to ICHEP 2000, Osaka, Japan
ZEUS Collaboration, Paper 1040 submitted to ICHEP 2000, Osaka, Japan

\bibitem{zrp}
ZEUS collaboration, abstract 1042, submitted to ICHEP00, Osaka

\bibitem{kuz01}
For a recent review and references, see\\
M.~Kuze, {\it Searches for new physics at HERA}, hep-ex/0106030

\bibitem{extrah1}
H1 Collaboration, C.~Adloff et al., Phys.~Lett. {\bf B 479} (2000) 358;\\
H1 Collaboration, Abstract 951 submitted to ICHEP00, Osaka, Japan

\bibitem{psd96}
E.~Perez, Y.~Sirois, H.~Dreiner, Proc.~{\it Future Physics at HERA}, 
Hamburg 1996, eds. G.~Ingelman, A.~De Roeck, R.~Klanner, p. 297

\bibitem{bw01}
W.~Buchm\"uller, R.~R\"uckl, in preparation

\bibitem{glt01}
For a recent discussion and references, see\\
A.~de~Gouv\^ea, S.~Lola, K.~Tobe, Phys.~Rev.~{\bf D 63} (2001) 035004

\bibitem{h1rp}
H1 Collaboration, C.~Adloff et al., Eur.~Phys.~J. {\bf C 20} (2001) 639

\bibitem{fus01}
T.Fusayasu, Moriond Conf. on {\it Electroweak Interactions}, Les Arcs, 2001

\bibitem{rpp00}
Review of Particle Physics, Eur.~Phys.~J. {\bf C 15} (2000) 1

\bibitem{hai01}
J.~Bl\"umlein, M.~Klein, T.~Riemann, in Proc. {\it The HERA Workshop}, 
Vol.~II, Hamburg 1987, ed. R.~D.~Peccei, p.687;\\
D.~Haidt, private communication

\bibitem{rs01}
A.~Ringwald, F.~Schrempp, Phys.~Lett. {\bf B 503} (2001) 331

\bibitem{h1zrise}
H1 Collaboration, I.~Abt et al., Nucl.~Phys. {\bf B 407} (1993) 515;\\
ZEUS Collaboration, M.~Derrick et al., Phys.~Lett. {\bf B 316} (1993) 412

\bibitem{h1zdiff}   
ZEUS Collaboratin, M.~Derrick et al., Phys.~Lett. {\bf B 315} (1993) 481;\\
H1 Collaboration, T.~Ahmed et al., Nucl.~Phys. {\bf B 429} (1994) 477

\bibitem{rx74}
A.~de~Rujula et al., Phys.~Rev.~{\bf D 10} (1974) 1649

\bibitem{dglap} 
V.~N.~Gribov, L.~N.~Lipatov, Sov.~J.~Nucl.~Phys. {\bf 15} (1972) 438, 675;\\
G.~Altarelli, G.~Parisi, Nucl.~Phys. {\bf B 126} (1977) 298;\\
Yu.~L.~Dokshitzer, Sov.~Phys.~JETP {\bf 46} (1977) 641 

\bibitem{grv93}  
M.~Gl\"uck, E.~Reya, A.~Vogt, Phys.~Lett.~{\bf 306 B} (1993) 391

\bibitem{bfkl}
L.~N.~Lipatov, Sov.~J.~Nucl.~Phys. {\bf 23} (1976) 338;\\
V.~S.~Fadin, E.~A.~Kuraev, L.~N.~Lipatov, Phys.~Lett. {\bf 60 B} (1975) 50;
Sov.~Phys.~JETP {\bf 44} (1976) 443; {\it ibid.} {\bf 45} (1977) 199;\\
Y.~Y.~Balitski, L.~N.~Lipatov, Sov.~J.~Nucl.~Phys. {\bf 28} (1978) 822

\bibitem{grv83}
L.~V.~Gribov, E.~M.~Levin, M.~G.~Ryskin, Phys.~Rep.~{\bf 100} (1993) 481;\\
for a review and references, see\\
A.~H.~Mueller, Proc.~{\it QCD: Perturbative or nonperturbative?}, 
Autumn School Lisbon 1999, hep-ph/9911289

\bibitem{abf01}
For a recent discussion and references, see\\
G.~Altarelli, R.~Ball, S.~Forte, Nucl.~Phys. {\bf B 599} (2001) 383

\bibitem{wal01}
H1 Collaboration, R.~Wallny, these proceedings

\bibitem{nag01}
ZEUS Collaboration, K.~Nagano, these proceedings

\bibitem{alh1}
H1 Collaboration, C.~Adloff et al., Eur.~Phys.~J. {\bf C 21} (2001) 33

\bibitem{alz}
ZEUS Collaboration, A.~Cooper-Sarkar, these proceedings

\bibitem{dl87}
A.~Donnachie, P.~V.~Landshoff, Phys.~Lett. {\bf B 191} (1987) 309

\bibitem{tv94}   
L.~Trentadue, G.~Veneziano, Phys.~Lett. {\bf B 323} (1994) 201

\bibitem{bs94}  
A.~Berera, D.~E.~Soper, Phys.~Rev. {\bf D 50} (1994) 4328

\bibitem{col98}
J.~C.~Collins, Phys. Rev. {\bf D 57} (1998) 3051

\bibitem{is85}   
G.~Ingelman, P.~E.~Schlein, Phys.~Lett. {\bf B 152} (1985) 256

\bibitem{gs96}
T.~Gehrmann, W.~J.~Stirling, Z.~Phys. {\bf C 70} (1996) 89

\bibitem{h101}
H1 Collaboration, C.~Adloff et al., Eur.~Phys.~J. {\bf C 20} (2001) 29

\bibitem{bex99}  
J.~Bartels, J.~Ellis, H.~Kowalski, M.~W\"usthoff, Eur.~Phys.~J. 
{\bf C 7} (1999) 443

\bibitem{ht01}
A.~Hebecker, T.~Teubner, Phys.~Lett. {\bf B 498} (2001) 16

\bibitem{gbw01}
K.~Golec-Biernat, M.~W\"usthoff, Eur.~Phys.~J. {\bf C 20} (2001) 313

\bibitem{bjo72}   
J.~D. Bjorken, AIP Conf.~Proc. No.6, eds. M.~Bander et al.
(AIP, New York, 1972) p. 151;\\
J.~D. Bjorken, J.~B.~Kogut, Phys.~Rev. {\bf D 8} (1973) 1341

\bibitem{bh96}   
W.~Buchm\"uller, A.~Hebecker, Nucl.~Phys. {\bf B 476} (1996) 203;\\
W.~Buchm\"uller, Phys.~Lett. {\bf B 353} (1995) 335

\bibitem{mlv94}    
L.~McLerran, R.~Venugopalan, Phys.~Rev. {\bf D 49} (1994) 2233

\bibitem{hw98}    
A.~Hebecker, H.~Weigert, Phys.~Lett. {\bf B 432} (1998) 215

\bibitem{nz91}
N.~N.~Nikolaev, B.~G.~Zakharov, Z.~Phys. {\bf C 49} (1991) 607

\bibitem{bgh99}
W.~Buchm\"uller, T.~Gehrmann, A.~Hebecker, Nucl.~Phys. {\bf B 537} (1999) 477

\bibitem{mue99}
A.~H.~Mueller, Nucl.~Phys. {\bf B 558} (1999) 285

\bibitem{gbw99}    
K.~Golec-Biernat, M.~W\"usthoff, Phys.~Rev. {\bf D 59} (1999) 014017; 
{\it ibid.} Phys.~Rev. {\bf D 60} (1999) 114023

\bibitem{zeus99} 
ZEUS Collaboration, J.~Breitweg et al., Eur.~Phys.~J. {\bf C 6} (1999) 43

\bibitem{kwx97}
J.~Jalilian~Marian, A.~Kovner, A.~Leonidov, H.~Weigert, Nucl.~Phys.
{\bf B 504} (1997) 415;\\
A.~Kovner, J.~G.~Milhano, H.~Weigert, Phys.~Rev. {\bf D 62} (2000) 114005

\end{thebibliography}
\end{document}